\documentclass{article}
\usepackage{amsmath,amsthm,amssymb}
\usepackage{graphicx}
\usepackage{url}

\newtheorem{prop}{Proposition}
\newtheorem{lemma}{Lemma}
\newtheorem{definition}{Definition}
\newtheorem{fact}{Fact}
\DeclareMathOperator{\var}{var}

\begin{document}
\title{Management Fads, Pedagogies, and other Soft Technologies}
\author{Jonathan Bendor\\
\small{Graduate School of Business, Stanford University, Stanford,
CA}
94305\\\\
Bernardo A. Huberman and Fang Wu\\
\small{HP Labs, Palo Alto, CA 94304}} \maketitle

\begin{abstract}
We present a model for the diffusion of management fads and other
technologies which lack clear objective evidence about their merits.
The choices made by non-Bayesian adopters reflect both their own
evaluations and the social influence of their peers. We show, both
analytically and computationally, that the dynamics lead to outcomes
that appear to be deterministic in spite of being governed by a
stochastic process. In other words, when the objective evidence
about a technology is weak, the evolution of this process quickly
settles down to a fraction of adopters that is not predetermined.
When the objective evidence is strong, the proportion of adopters is
determined by the quality of the evidence and the adopters'
competence.
\end{abstract}
\pagebreak
\section{Introduction}

In domains such as management and education, organizational
practices often seem to come and go in puzzling ways.  They are
introduced with fanfare, but then they diffuse with little evidence
that they work well. While they are sometimes discarded at later
times, they are so with little conclusive evidence about their
performance. Consider, for example, Quality Circles. Possibly as
many as 90 percent of the Fortune 500 companies had adopted QCs by
1985, but by 1987 ``more than 80\% of the \emph{Fortune} 500
companies that adopted QCs in the 1980s had abandoned them by 1987''
\cite[pp.~256]{abrahamson:96}. Yet one is hard pressed to find hard
evidence of their impact, even after the fact.

Fads are so common in American education that one observer reports
that ``School leaders in a whimsical mood sometimes play a parlor
game called `Spot That Jargon,' in which the goal is to name as many
past educational fads as possible.  The list is usually
impressive.'' \cite{lashway:98, maddux:04}

This paper examines the diffusion of such innovations or ideas.  We
call them \emph{soft technologies}, not because of their physical
properties but because evidence for or against them is equivocal,
inconclusive, or even nonexistent.  We contend that the choices made
by adopters are quasi-rational: they reflect both an attempt to
assess the imperfect data surrounding such innovations as well as a
reliance on social cues, i.e., what peers have done.  We argue that
these two elements are linked by what could be called Festinger's
Hypothesis: the more equivocal the evidence, the more people rely on
social cues \cite[pp.~118]{festinger:54}.

In this paper we present a model that considers soft technologies as
those for which (1) the objective evidence is weak and (2) people
rely heavily on the prior choices of people in similar roles. We
then show that the dynamics of the model leads to outcomes that
appear to be deterministic in spite of being governed by a
stochastic process. In other words, when the objective evidence for
the adoption of a soft-technology is weak, any sample path of this
process quickly settles down to a fraction of adopters that is not
predetermined by the initial conditions: ex ante, every outcome is
just as (un)likely as every other. In the case when the objective
evidence is strong, the process settles down to a value that is
determined by the quality of the evidence. In both cases the
proportion of adopters of the technology never settles into either
zero or one.

\subsection{Related Work}

In the most highly developed mathematical models of fads---economic
theories of ``herding''---decision makers also use social cues but
do so in perfectly rational ways, via Bayesian
updating.\footnote{There is now a large literature on informational
cascades triggered by fully rational agents: see the annotated
bibliography of Bikhchandani, Hirshleifer and Welch, available on
the Web \cite{bikhchandani:96}. For seminal papers in this line of
work see \cite{banerjee:92, bikhchandani:92, welch:92}.}  Though we
agree that social cues matter we think that the premise of
Bayesianism exaggerates the rationality of agents facing the
difficult decision of whether or not to adopt a soft technology. In
particular, there is little evidence for the claim that people are
perfect Bayesians. A leading experimental economist summarizes the
evidence as follows:
\begin{quote}
Much research in cognitive psychology suggests that the way in which
people form judgments of probability departs systematically from the
laws of statistics and from Bayesian updating. (This should not be
surprising, because there is no reason to think that evolution of
brain processes like memory, language, perception, categorization,
and reasoning would have adapted us to use a rule that Bayes only
``discovered'' a couple of hundred years ago.)  Some research points
toward systematic departures, or ``biases'', which spring from a
small number of ``heuristics'', like anchoring, availability, and
representativeness \cite[pp.~171]{camerer:98}.
\end{quote}

Thus, the \emph{theoretical} value of herding models---the
intriguing demonstration that what appears to be conformity behavior
in the aggregate is consistent with perfectly rational action of
individuals---should not be confused with \emph{empirical}
confirmation of its micro-postulates.  As a purely theoretical point
it is interesting to recognize that perfect information-processing
by individual agents is, under certain circumstances, consistent
with conformity-like behavior. But we suspect that to the extent
that such models receive empirical support, the support will be
``weak'' in the sense that the data on conformity or herding will
also be consistent with a wide variety of other
sensible-though-suboptimal forms of individual
information-processing. Again, Camerer's assessment of Bayesianism
is pertinent:
\begin{quote}
As a descriptive theory, Bayesian updating is weakly grounded in the
sense that there is little direct evidence for Bayesian updating
which is not also consistent with much simpler theories. Most of the
evidence in favor of Bayesian updating boils down to the fact that
if new information favors hypothesis A over B, then the judged
probability of A, relative to B, rises when the information is
incorporated.  This kind of monotonicity is consistent with Bayesian
updating but also with a very wide class of non-Bayesian rules (such
as anchoring on a prior and adjusting probabilities up or down in
light of the information) \cite[pp.~171]{camerer:98}.
\end{quote}

In this paper we propose a model that is consistent with all of
Camerer's observations and so is an alternative to canonical herding
models. Thus our agents exhibit normatively desirable and
empirically plausible monotonicity properties: in particular, the
more the social cues favor innovation A over B, the more likely it
is that an agent will select A, ceteris paribus.  Yet the reasoning
that underlies such choices is adaptively rational rather than fully
rational. Moreover, unlike many adaptive models of fads, the present
model generates analytical solutions, not just computational
ones.\footnote{Many---perhaps most---adaptive models of fads are
what has come to be called ``agent-based models'' and it is
virtually a defining feature of such models that they be
computational.  (For a survey of agent-based models, including
several applied to fads, see \cite{macy:02}.)}

\section{The Model}

A world in which people can make mistakes (e.g., adopt an inferior
method of instruction, partly because many other school districts
have already done so) and where they are influenced by the possibly
erroneous, possibly correct choices of similar decision makers is
inherently probabilistic.  Hence our model is built around a
probabilistic choice process.

We assume that two alternatives, A and B, diffuse through a
population of decision makers.\footnote{We shall often interpret A
and B as competing innovations, but the model allows for different
interpretations.  For example, one of the options could be the
population's status quo alternative while the other is an
innovation.  We will return to this specific interpretation in
Section \ref{sec:interpretations}.} In every period one decision
maker makes up his mind about whether to adopt A or B; this choice
is final. (In this sense the formulation is like most ``contagion''
models.) The diffusion continues until everyone in the population
has selected either A or B.  To get the process going, we assume
that initially ($t=0$) at least one person has made a choice, i.e.,
at least one person already champions either A or B.  However, we
allow for the possibility that either option may have multiple
initial champions. (One could regard these early champions as the
inventors of the two alternatives.) A useful benchmark case to keep
in mind is a fair start, in which A and B are backed by the same
number of initial champions. The numbers of A- and B-champions at
$t=0$ will be denoted by $m_0$ and $n_0$.

The heart of the model is how agents decide on which option to
adopt. As noted earlier, we assume that there are two components to
the adoption decision.  The first is based on individual judgment;
the second, on social influence.  Regarding the first component, we
assume that a person isolated from social influence would choose the
objectively superior option (labeled A in our model) with
probability $p$. Thus $p$ reflects the quality of the evidence about
the relative merits of A versus B, plus whatever a priori bias
(possibly due to a folk theory) exists.  If A's superiority is
obvious then $p$ will be close to 1; if the two alternatives are
nearly interchangeable or if evaluation technologies are primitive
then $p$ will be close to $1/2$.  If there is an a priori,
theory-driven bias against A, then $p$ could be less than $1/2$. In
general $p$ is in $(0,1)$.

We assume that the impact of social influence is linearly increasing
in the proportion of the ``converted'' who have adopted in a
particular adoption.  Thus, if $M_t$ denotes the number of people
who by period $t$ have chosen A and $N_t$ denotes the number who
have selected B, then the social pressure to choose A is simply
$M_t/(M_t+N_t)$. The social pressure to choose B is, of course,
$N_t/(M_t+N_t)$.\footnote{This is consistent with a simple search
process: if the decision maker looks for social cues (i.e., the
choices that the already-converted have made), then with probability
$M_t/(M_t+N_t)$ the first convert she bumps into is an A-adherent.
And so with that probability she is persuaded to adopt A
(conditional on her choosing via social imitation).}  (We will use $m_0$ and
$n_0$ to denote the initial number of adherents to A and B, respectively.)

Since we are trying to construct a simple benchmark model that has a
clean structure, we assume that an agent's choice is simply a
weighted average of the above components (individual judgement and
social influence). Thus
\begin{equation}
\label{eq:dyn-A}P[\text{agent in period $t+1$ chooses }A] = \alpha
\cdot p + (1-\alpha) \frac{M_t}{M_t+N_t}
\end{equation}
where $0 \le \alpha \le 1$.  Similarly, the probability that the
chooser in period $t+1$ selects B is
\begin{equation}
P[\text{agent in period $t+1$ chooses }B] = \alpha \cdot (1-p) +
(1-\alpha) \frac{N_t}{M_t+N_t}.
\end{equation}

Festinger's hypothesis---that people are more open to social
influence when evidence is equivocal---amounts here to assuming that
$\alpha$ and $p$ are positively correlated if $p>1/2$ and negatively
correlated if $p<1/2$. We believe that this is a sensible
proposition but we do not require it for our analytical results.  We
do use it in most of our simulations, however.  Moreover,
Festinger's hypothesis informs our understanding of what we consider
soft technologies: in the context of Eq.~(\ref{eq:dyn-A}), a soft
technology is one with a $p$ in the vicinity of $1/2$ and a low
$\alpha$.

To get a feel for how this process works it is useful to consider
first the two extreme cases: i.e., when $\alpha=1$ and $\alpha=0$.
The former is simply an independent trials process with a
probability of ``heads'' of $p$. This process and its properties are
well-understood. The case of $\alpha=0$ (pure social influence) is
the standard Polya's urn process \cite{polya:23}. Given that we are
particularly interested in soft technologies, which here are
represented by low values of $\alpha$, we will pause for a moment to
recapitulate its features.

\section{Pure social influence ($\alpha=0$)}

Suppose 100 people have made up their minds, with 70 having chosen A
and 30 having chosen B.  Then in the current period agent 101 has a
70\% chance of selecting A.  Consequently the \emph{expected}
proportion of the population who cleave to A is $.7 (\frac{71}{101})
+ .3 (\frac{70}{101})=\frac 1{101} (.7\cdot 71 + .3\cdot 70)=\frac
1{101} (70.7)=.70$; i.e., the expected proportion exactly equals the
current proportion.  It is easy to show that this \emph{martingale
property} holds in general: on average the pure social influence
process stays exactly where it currently is. That is,
\begin{equation}
\label{eq:martingale}E\left[F_{t+k} \Big|\, F_t= \frac{M_t}{M_t+N_t}
\right] = \frac{M_t}{M_t+N_t} = F_t
\end{equation}
for all $k>0$. Hence the pure social influence process is strongly
path dependent: in expectation it tends to stay wherever it
is---i.e., wherever it has arrived via the particular sample path it
has been traveling.

\begin{prop}
\label{prop:polya}If $m_0=n_0=1$ (and $\alpha=0$), then $P[M_t=1]=
P[M_t=2] = \cdots = P[M_t=t+1] = 1/(t+1)$, for all $t>0$.
\end{prop}
\begin{proof}
This is a classical result \cite{polya:23}.
\end{proof}

Thus every feasible outcome is equally likely at every date: given a
fair start, the pure process of linear social pressure is completely
`blind'. Therefore the process is just as likely to wind up
generating a heterogeneous diffusion, with half the population
championing A and the other half B, as it is to wind up with a
sharply skewed outcome, with nearly everyone backing A.\footnote{It
is worth mentioning that Proposition \ref{prop:polya} depends upon
the initial seed being $m_0=n_0=1$. If it is a fair start but there
are more two initial champions then the resulting distributions are
not uniform, though they are symmetric around $1/2$. Nevertheless,
the martingale conditional expectations property \emph{does}
continue to hold, for any values of $m_0$ and $n_0$. So in this
sense the strong path dependence property is insensitive to the
initial conditions.}

\begin{figure}
\begin{center}
\includegraphics[scale=.7]{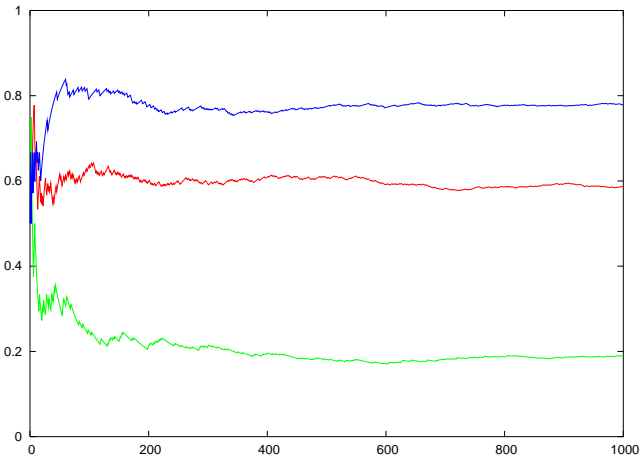}
\caption{\label{fig:polya}A simulation of the pure Polya's urn
process for 1000 rounds. The initial condition is one A and one B.
As can be seen from the figure, the three sample paths converge to
different limits.}
\end{center}
\end{figure}

It is instructive to look at a typical sample path; see
Fig.~\ref{fig:polya}. Note that initially the relative frequencies
of A- versus B-adherents swing wildly: with only a small number of
initial converts, early adopters have a lot of influence. But once
hundreds of people have taken sides subsequent adopters have little
impact on relative frequencies, and so the process settles down.
\emph{Hence it may appear to agents inside the process that the
diffusion is moving toward some predetermined equilibrium.} Part of
this impression is correct: as $t \to \infty$, any sample path of
this process will settle down near some long-run proportion of
A-adherents and B-adherents.  But we know from Proposition
\ref{prop:polya} that this asymptotic state was not at all
predetermined: ex ante, every outcome is just as (un)likely as every
other.

What does the pure Polya process tell us about the complex process,
in which objective evidence does play a role? Recall that by
Festinger's hypothesis, soft technologies have low $\alpha$'s.
Moreover, an agent's adoption probability, as represented by
Eq.~(\ref{eq:dyn-A}), is continuous in $\alpha$. Hence for $\alpha$
``close'' to zero the diffusion will approximate a pure social
influence process: it will be ``nearly'' blind, ex ante.

In such circumstances agents are influenced by a mixture of
conformity and individual judgment about the evidence.  Yet they may
not have good access to the fine structure of their own choice
processes \cite{wegner:98}.  For one thing, they may not recognize
how much of their choice is influenced by social cues. And even if
they do, they may rationalize that part of it, along the lines of
herding models: the behavior of peers conveys information, and so it
is rational to be socially influenced.  And of course that may be
so. But if $\alpha$ is sufficiently low there is a good chance (less
than half but still appreciable) that at date $t$ a majority of the
converts will back the weaker option; hence the next adopter could
be led astray.

Further, we suspect that people involved in a diffusion of a soft
technology do not have good intuitions for the stochastic properties
of the process.  In particular, we suspect that the fact that with a
low $\alpha$ a soft technology could, given a large population of
adopters, wind up at many different outcomes with nearly equal ex
ante probabilities is underappreciated.  Life unfolds as a sample
path, and the ``settling down'' feature of the particular sample
path one inhabits (as in Fig.~1) will be much more salient than
theoretical ex ante probabilities---if the latter are recognized at
all.

\section{Properties of the Complex Process: $\alpha \in (0,1)$}

\subsection{The mean of the process}

We now directly investigate properties of the complex process.
First let us examine how it behaves over time.

\begin{prop}
\label{prop:EF_monotone} Assume $\alpha>0$. $EF_t\to p$ monotonously
as $t\to \infty$.
\end{prop}

In fact, instead of expectations, we can establish a strong
convergence result:

\begin{prop}
\label{prop:strong_conv}$F_t\to p$ a.s. as $t\to\infty$.
\end{prop}

\begin{figure}
\begin{center}
\begin{minipage}{10cm}
\centering
\includegraphics[scale=.7]{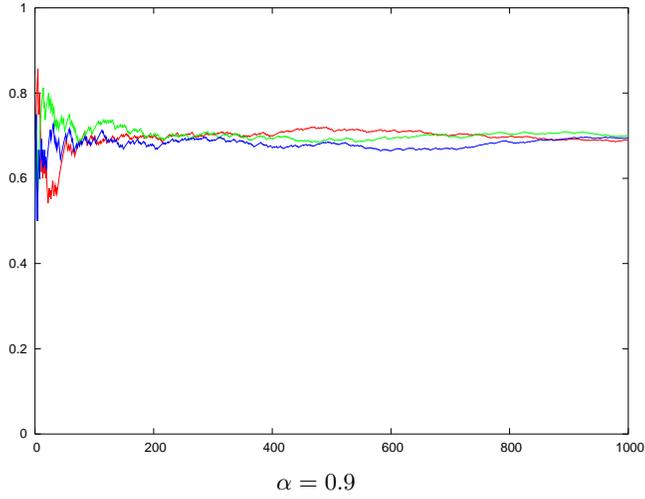}\\
\small{$\alpha=0.9$}
\end{minipage}

\bigskip

\begin{minipage}{10cm}
\centering
\includegraphics[scale=.7]{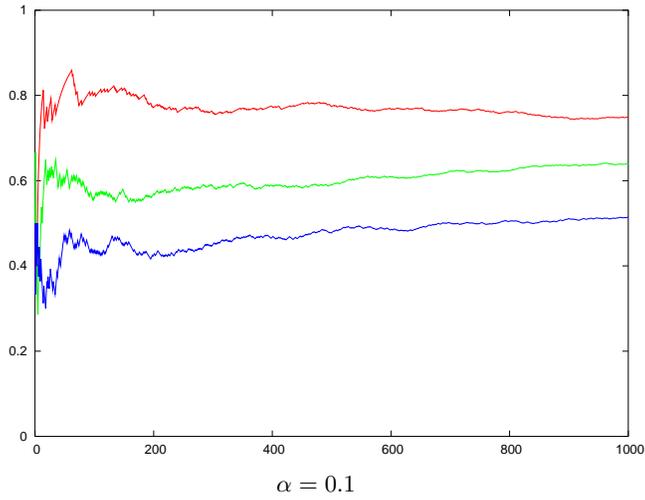}\\
\small{$\alpha=0.1$}
\end{minipage}
\caption{Simulations of the complex process after $1000$ rounds. In
both cases ($\alpha=0.9$ and $0.1$) we choose $p=0.7$ and initial
condition one A and one B. In each figure three sample paths are
shown. As can be seen, the convergence rate for small $\alpha$ is
much slower than the rate for large $\alpha$.}
\end{center}
\end{figure}

Thus the process exhibits a constant drift or bias toward $p$, the
value of individual judgment. This reason behind this is because the
martingale property is broken. In fact, instead of
Eq.~(\ref{eq:martingale}), it can be shown (see Appendix) that for
the complex process $\alpha\in (0,1)$,
\begin{equation}
\label{eq:mean_dynamics}E[F_{t+1}|F_t] = F_t +
\frac{\alpha(p-F_t)}{t+s+1}.
\end{equation}

Note that the process could be mostly one of social
construction---most of the weight is on imitation---yet improvement
will tend to occur anyway, if $\frac{m_0}{m_0 + n_0}< p$. Henceforth
we denote $\frac{m_0}{m_0 + n_0}$ by $f_0$.

Next let us consider how $F_t$ is affected by variations in the
model's two basic parameters, $\alpha$ and $p$.  The latter's effect
is both obvious and unconditional.  Clearly, the average fraction of
correct choices, $EF_t$, is increasing in $p$.  But more can be
said.

\begin{prop}
\label{prop:high_p}A higher value of $p$ yields a distribution of
$F_t$ that stochastically dominates a distribution of $F_t$ produced
by a lower value of $p$, for all $t>0$.
\end{prop}

This immediately implies that $EF_t$ is increasing in
$p$.\footnote{However, the converse does not hold: an increase in
the mean fraction of correct choices does not imply that one
distribution stochastically dominates the other.  So Proposition
\ref{prop:high_p} is stronger than just a comparison of means.}

It is worth emphasizing that this strong effect obtains even if the
process is mostly one of social construction (low $\alpha$'s).  Thus
the presence of social construction does \emph{not} utterly prevent
``rational engineering'' (e.g., concerning the technology of
evaluation) from having benign effects.  (Of course, in a sense this
is built into the model, but hopefully it is built-in in a plausible
way.)

Further, Festinger's hypothesis implies that if $p > 1/2$, then as
the evaluation technology improves, decision makers will rely more
on the evidence and less on social cues: the rise in $p$ will be
followed by an increase in $\alpha$.  This \emph{indirect} impact of
changes of $p$ might be just as important as its direct effect on
$EF_t$.  This naturally raises the question, how do changes in
$\alpha$ affect the process?

\begin{prop}
\label{prop:high_alpha}\mbox{}
\begin{itemize}
\item[(i)] If $f_0<p$, then $EF_t$ is increasing in $\alpha$, for all
$t>0$.
\item[(ii)] If $f_0>p$, then $EF_t$ is decreasing in $\alpha$, for all
$t>0$.
\end{itemize}
\end{prop}

Thus, absent a lucky start ($f_0>p$), less reliance on social cues
improves matters.  Consequently, if Festinger's hypothesis is
correct, improvements in evaluation technology have \emph{two}
benign effects on diffusions that don't enjoy lucky starts: the
obvious direct one (Proposition \ref{prop:high_p}) and a less
obvious indirect one (Proposition \ref{prop:high_alpha}), via
increases in $\alpha$.

\subsection{The unpredictability of the process}

If we compare the two pure processes, it is evident that, given a
standard start ($m_0=n_0=1$), the pure social process is less
predictable (as measured by the variance of $F_t$).  When the
diffusion is purely a matter of social construction, every feasible
outcome is equally likely; this is intuitively very random, and it
translates into one with a high variance.  In contrast, even the
variant of the pure individualistic process with maximal variance
($p=1/2$), extreme outcomes---nearly everyone choosing A or B, for
example---are less likely than other outcomes.  So the variance must
be less than that produced by pure social imitation.

Since the complex process is, at the level of individual choice, a
weighted average of the two pure processes, it is intuitively
reasonable that the variance of the complex process would be
increasing in $1-\alpha$, the weight on the higher variance
component, given a fair start. This intuition is correct, but since
it conditions on a particular starting point it is incomplete. The
next result shows that more weight on the social component can
\emph{reduce} the variance of the complex process, at least for
awhile.

\begin{prop}
\label{prop:alpha} $\partial(\var F_t)/\partial \alpha <0$ for all
$t>0$ iff either
\begin{itemize}
\item[(i)] $f_0<p$ and $p_0=(1-\alpha)f_0+\alpha p>1/2$, or
\item[(ii)] $f_0>p$ and $p_0=(1-\alpha)f_0+\alpha p<1/2$.
\end{itemize}
\end{prop}

Thus for parameters in this range, increases in the weight on
evidence both improve community-wide accuracy (on average) and
reduce the variance of outcomes.

While it makes sense that the more socially constructed a diffusion
is the more variable are its outcomes, it is important to emphasize
that this conclusion does \emph{not} always hold, as is indicated by
the ``only if'' part of Proposition \ref{prop:alpha}.   Here is why.
Suppose initially A-adherents are few and far between; for
simplicity, let $f_0=0$.  Then if diffusion were based only on
imitation ($\alpha=0$), everyone would adopt the wrong option---and
the process, although built completely on social cues, would exhibit
no variability whatsoever.  In such a case small increases in
$\alpha$ would \emph{increase} the variance of the outcomes, at
least in the early goings. Of course this increase in variance is
thoroughly benign: without it, the diffusion would be stuck in a
highly predictable but consistently inferior sample path.

\begin{prop}
\mbox{}
\begin{itemize}
\item[(i)] $\partial(\var F_t)/\partial p<0$ for all $t>0$ if $p\ge
1/2$ and $p_0>1/2$;
\item[(ii)] $\partial(\var F_t)/\partial p>0$ for all $t>0$ if $p\le
1/2$ and $p_0<1/2$.
\end{itemize}
\end{prop}

Regarding part (ii), note that increases in $p$ increase the
variance in a socially desirable way.  However, people who like
predictability could resent the ``rationalizing'' effect of
improvements in program information and evaluation technology,
because in these circumstances increasing $p$ makes things
``messier'' and less predictable.

As a matter of interpretation, one might say that a $p\le 1/2$ and
an $p_0<1/2$ involves a ``soft'' technology with a vengeance: there
is some kind of bias against the better technology, maybe because it
is not as trendy, and there is also a bias in terms of the initial
social proclivities ($p_0<1/2$). In this case, improving evaluation
technology also creates a ``messier'' and more confusing process ex
ante, in that such improvements increase the variance of the
outcomes.

Indeed, even if $p>1/2$ (i.e., the ``standard'' case), we know that
in period 1 the variance of $F_1$ is increasing in $p$ if $p_0<1/2$.
(And probably this will hold for some finite number of periods after
period 1 too.)  This is a very natural case of a soft technology:
there is a social bias against the better technology (bad luck,
essentially, or maybe glamour is on the side of the weaker
technology), in that $f_0<1/2$, and things are noisy enough so that
$p$, though above $1/2$, is still fairly low. Hence if Festinger's
hypothesis were to kick in, so that $\alpha$ were pretty low too,
then $p_0<1/2$, and the variance of $F_1$ would be increasing in
$p$.

\subsection{The effects of initial conditions}

There are two types of initial conditions: the \emph{size} of the
initial seed ($m_0 + n_0$) and the \emph{proportion}
($f_0=\frac{m_0}{m_0+n_0}$). Each has effects independently of the
other (i.e., holding the other fixed), so we will examine them
separately and in a ceteris paribus manner.

\subsubsection{Varying the size of the initial seed}

\begin{prop}
Suppose $F_t'$ is a bigger process than $F_t$ in that $m_0'=km_0$
and $n_0'=kn_0$, where $k$ is an integer larger than one. In all
other respects the two processes are identical.
\begin{itemize}
\item[(i)] $EF_t'<EF_t$ for all $t>0$ iff $f_0<p$.
\item[(ii)] $EF_t'>EF_t$ for all $t>0$ iff $f_0>p$.
\end{itemize}
\end{prop}

Thus a bigger initial seed acts as an inertial anchor, slowing down
the movement of $E[F_t]$ to its attractor $p$.

\subsubsection{Varying the initial proportions}

It is obvious that, all else equal, increasing the initial bias
toward A (i.e., increasing $f_0$) boosts $EF_t$ at every date.
Suppose, e.g., we start out with $m_0+1$ A-converts, instead of with
just $m_0$.  This increases the social conformity pressure for the
agent in period 1 to adopt A.  That in turn increases $F_1$, the
expected fraction of the converts who adhere to A at the end of
period 1, which in turn provides more social cues to the decision
maker in period 2 to adopt A and so on.

But a higher $f_0$ has an even stronger effect---stronger even than
ordinary first-order stochastic dominance.  To see what this sense
is, consider the following definition.

\begin{definition}
Suppose the support of $F_t$ and $F'_t$ can be divided into three
mutually disjoint subsets: a non-empty set of high states $\{
f_{h_1,t}, f_{h_2,t}, \dots, f_{h_a,t} \}$, a non-empty set of low
states $\{ f_{l_1,t}, f_{l_2,t}, \dots, f_{l_b,t} \}$, and a
possibly empty set of intermediate states $\{ f_{m_1,t}, f_{m_2,t},
\dots, f_{m_c,t}\}$. Suppose the three subsets are connected in the
sense that any low state is less than any intermediate state, and
any intermediate state is less than any high state. We say that
$F'_t$ is stochastically bigger than $F_t$ in a strong sense if
$P[F'_t = f_{h,t}] > P[F_t = f_{h,t}]$ for any high state $h$,
$P[F'_t = f_{m,t}] = P[F_t = f_{m,t}]$ for any intermediate state
$m$, and $P[F'_t = f_{l,t}] < P[F_t = f_{l,t}]$ for any low state
$l$.
\end{definition}

Thus a distribution of $F_t$ that is stochastically bigger in a
strong sense puts strictly more weight on high fractions of converts
who have chosen correctly, and strictly less weight on low
fractions. (Clearly this implies first-order stochastic dominance,
but the converse need not hold, so this is in fact a ``strong
sense'' of stochastic dominance.)

\begin{prop}
All else equal, the distribution of $F_t$ is stochastically
increasing, in a strong sense, in $f_0$, for all $t>0$.
\end{prop}

Naturally, smaller $f_0$'s produce distributions of $F_t$ that are
stochastically smaller, in a strong sense.  Suppose, therefore, that
A is an innovation that is superior to B.  If B is not an
innovation---it is in fact the status quo---and the community is
quite traditional (everyone initially uses B, though people are
willing to consider an innovation), then effectively $f_0$ equals
zero.  This is the toughest possible starting point for an
objectively superior innovation.

The effect of varying initial bias on the variance of the process is
perhaps less intuitive.

\begin{prop}
$\partial (\var F_t)/\partial f_0 <0$ for all $t>0$ if $p_0>1/2$ and
$p>1/2$.
\end{prop}

\subsection{The convergence rate}

If we wish to use our complex process to explain diffusion of
innovations in the real world, only studying the limit behavior is
not enough. If the characteristic time needed to reach the asymptote
state is too long to be observed, then the asymptote state cannot be
very meaningful in the practical sense. A rough estimation of the
convergence rate of the complex process can be achieved from a
mean-field approximation. Neglecting the noise term, the stochastic
process $F_t$ can be approximated by Eq.~(\ref{eq:mean_dynamics}):
\begin{equation}
F_{t+1}=F_t + \frac{\alpha(p-F_t)}{t+s+1}.
\end{equation}
Further approximating $F_t$ by a continuous process $F(t)$, we can write
\begin{equation}
\frac{dF(t)}{dt} = \alpha \frac{p-F(t)}{t+s+1} \sim \alpha \frac{p-F(t)}t.
\end{equation}
Solving for $F(t)$, we find
\begin{equation}
|F(t)-p| \sim t^{-\alpha}.
\end{equation}
If we define a characteristic convergence time $T$ to be the time it takes
 $F(t)$ to converge to a vicinity within $\epsilon$ from $p$, then we have
\begin{equation}
T \sim \left(\frac 1\epsilon\right)^{\frac 1\alpha}.
\end{equation}
Thus as $\alpha\to 0$, the characteristic convergence time diverges
exponentially in $1/\alpha$. Indeed, when $\alpha=0$ (Polya's urn)
the process \emph{never} converges to $p$. (To be precise, the Polya
process can converge to any $p\in [0,1]$, but the probability that
it converges to any particular $p$ is zero.)

The mean-field estimation is in principle only for the mean of $F_t$, of course. For a fine estimation of the variance of $F_t$, we need a central limit theorem.

\begin{prop}
\mbox{}
\begin{itemize}
\item[(i)] If $1/2<\alpha\le 1$, then as $t\to \infty$, $\sqrt t(F_t-p)$
converges in distribution to a normal distribution with mean zero
and variance $p(1-p)/(2\alpha-1)$. In particular, $\var F_t =
O(1/t)$.
\item[(ii)] If $0< \alpha \le 1/2$, then $\var F_t \to 0$
slower than $O(1/t)$.
\end{itemize}
\end{prop}

\subsection{Implications regarding herding and massive conformity}

Assume, as in cascade models with rational agents, that the
``initial seed'' is exactly one person who chooses A with
probability $p$ and $B$ with probability $1-p$, and everyone else
follows sequentially.

\begin{definition}
We say that there is ``herding'' if after some period $T$ everyone
makes the same choice.\footnote{This is the term some scholars
(e.g., \cite{banerjee:92}) use.  Others (e.g.,
\cite{bikhchandani:96}) call this an informational cascade.}
\end{definition}

\begin{fact}
Herding occurs in our model (with positive probability) iff one of
the following condition holds:
\begin{itemize}
\item[(a)] $\alpha>0$ and $p=1$; then we get herding on option A.
\item[(b)] $\alpha>0$ and $p=0$; then we get herding on option B.
\item[(c)] $\alpha=0$; then we get herding on option A with
probability $p$ and herding on option B with the complementary
probability $1-p$.
\end{itemize}
\end{fact}
\begin{proof}
\emph{Sufficiency.} All three parts are trivial, by induction.

\medskip\noindent
\emph{Necessity.} Suppose none of (a)--(c) hold. Then we know that
$\alpha>0$ and $p\in (0,1)$. By Prop.~\ref{prop:strong_conv},
$F_t\to p$ almost surely. Hence both A and B are chosen infinitely
often with probability one.
\end{proof}

\begin{fact}
Under the assumptions of our model, if heterogeneous behavior ever
occurs then herding is impossible (with probability one).
\end{fact}
\begin{proof}
By Fact 1, if heterogeneous behavior has emerged by some date $t>0$
then the extreme conditions (a)--(c) in Fact 1 cannot hold. Hence we
know that $\alpha>0$ and $p\in (0,1)$. The rest is the same as the
necessity part of Fact 1.
\end{proof}

Conformity could be overwhelming---\emph{nearly} everyone in a
community winds up making the same choice---without being complete.
One of the surprising results of the rational cascade models is that
massive-conformity-in-the-making can be very fragile.  As
Bikhchandani, Hirshleifer and Welch put it, ``A little bit of public
information (or an unusual signal) can overturn a long-standing
\emph{informational cascade}.  That is, even though a million people
may have chosen one action, seemingly little information can induce
the next million people to choose the opposite action.  Fragility is
an integral component of the \emph{informational cascades} theory!''
(\cite{bikhchandani:96} original emphasis).

\begin{figure}
\centering
\includegraphics[scale=.7]{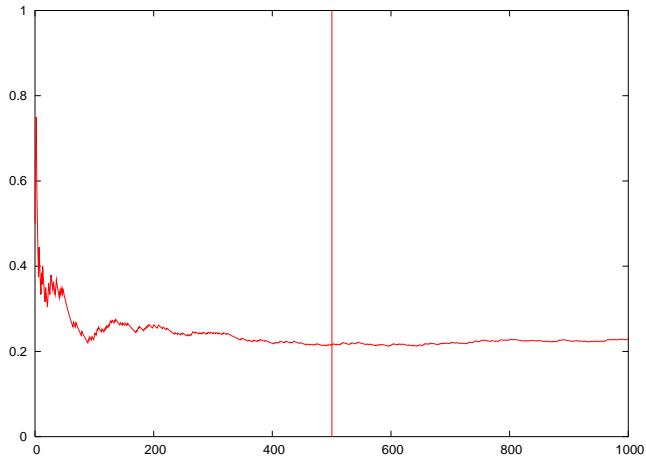}
\caption{\label{fig:change_p}A simulation of the complex process
with $\alpha=0.5$ and initial seed one A and one B. $p$ is set to
$0.2$ for the first 500 rounds and is updated to $0.3$ for another
500 rounds. The vertical line indicates the change point. As can be
seen, the sample path is not significantly affected by the new level
of $p$.}
\end{figure}

But this fragility is intimately linked to the agents' complete
rationality and deep understanding of informational cascades.  As
Bikchandani et al.~remark, in standard rational models ``everyone
knows that there is very little information in a cascade
\cite{bikhchandani:96}. That ``everyone'' pertains, of course, to
the model's rational agents, not to real decision makers:  the
evidence is that the latter do \emph{not} realize that there is very
little information in a cascade.\footnote{This is the pattern that
Kubler and Weiszacker found in their experiments on cascades
\cite{kubler:04}. As they put it, ``players do not consider what
their predecessors thought about their respective predecessors.
Thus, they do not understand that some of the decisions they observe
have been herding decisions, not based on any private information
(pp.~438).}  The agents in our model are imperfectly rational and
lack a deep understanding of cascades. Hence, diffusion processes in
our model are \emph{not} fragile (in the above sense). This is
easily established by re-inspecting Eq.~(\ref{eq:dyn-A}): the
probability that an agent makes the correct decision is continuous
in $p$, so a little bit of new public information---represented as a
sudden positive shock to $p$---will only increase the probability of
choosing correctly by a little bit. (For an illustration of this,
see Fig.~\ref{fig:change_p}.)

\section{\label{sec:interpretations}Alternative Interpretations}

It is worth noting that one can use the model to represent
situations in which there is a status quo option that everyone
already uses.  Then the diffusion is simply this: in every period
one agent has an opportunity to either take up the innovation (say,
A) or keep the status quo (say, B).  Further, one could allow for
the innovation to be objectively \emph{inferior} to the status quo.

Under this interpretation of a novel technology competing with a
status quo, it is reasonable to suppose that $f_0$ equals zero. This
is, from the point of view of an innovation, the toughest possible
starting point.

If it's a soft innovation then presumably both $p$ and $\alpha$ are
relatively low, so it is probably the case that $\alpha p +
(1-\alpha) f_0$ is less than one-half.  Hence by Proposition
\ref{prop:alpha}, decreasing the reliance on social cues (i.e., a
smaller $1-\alpha$) would \emph{increase} the variance of outcomes.

\section{Extensions of the Model}

\subsection{Endogenizing Festinger's Hypothesis}

If Festinger's hypothesis is correct, then the weight on social
cues, should depend on how difficult the choice is.  That is,
$\alpha$ should be a function of $p$.  We can stipulate a priori
several properties that this function should have.  First,
$1-\alpha$ should reach its maximum when $p=1/2$: it is in this
circumstance that an agent is maximally uncertain about the relative
merits of A versus B and so, following Festinger, s/he would be
maximally reliant on social cues.  Second, as a benchmark the
function should be symmetric around one half: $f(.5-k) = f(.5+k)$,
for $.5 \le k \le 1$.

A simple function with both properties is $1-\alpha = 4p(1-p)$. Then
the probability that the chooser in period $t$ would pick A would
equal $4p(1-p)\cdot F_t + 1-4p(1-p)\cdot p$.

An interesting feature of this choice equation is that it creates
the possibility that the probability of a correct choice is
\emph{decreasing} in $p$.  This seems bizarre but it is explicable:
it arises because $\alpha$ is a function of $p$.  Note that
$4p(1-p)$, the weight on social cues, is increasing when $p$ is less
than one half, since in this range increases in individual-level
accuracy make the choice problem more confusing or troubling.  Thus,
as we suggested earlier, increases in $p$ have two effects: a direct
effect on individualistic choice (which is always benign, as shown
by Proposition \ref{prop:high_p}) and an indirect one on the weights
(which is not always benign).  For certain parametric values the
indirect effect is sufficiently large and negative so as to swamp
the positive direct effect.\footnote{To see this concretely, note
that the derivative of $4p(1-p)\cdot f_0 + 1-4p(1-p)\cdot p$ with
respect to $p$ equals $4(1-2p)\cdot f_0 + (1-8p + 12p^2)$.  Then
suppose for example that $f_0 = 0$ and $p=.25$.  With these
parametric values the derivative equals $-.25$, so small increases
in $p$ make the agent in period one more likely to err.  However, if
$p$ is sufficiently big---specifically, if it exceeds $\max(.5,
f_0)$---then increases in $p$ do decrease the chance of mistakes.
(When $p>.5$ then a higher $p$ decreases the decision maker's
confusion, thus raising the weight on individualistic judgment,
which is benign since $p>f_0$.)}

Thus endogenizing Festinger's hypothesis will generate interesting
hypotheses.

\subsection{Diffusion in Communities with Internal Structure}

\subsubsection{Status Hierarchies}

There is considerable empirical support for the hypothesis that the
higher the decision maker's status, the more impact his/her adoption
choice has on the unconverted (e.g., \cite[pp.~275]{strang:98}). It
would be easy to incorporate this empirical regularity in a stark
way in the current model: only high status agents are imitated; the
adoption choices of low status agents are ignored.

An intriguing issue concerns the relative competence of high status
and low status agents.  In diffusions that tap into a relatively
strong scientific or technical base (pharmaceuticals, computer
hardware, etc.), one would expect higher status agents to be more
likely to make the right choice (higher $p$'s).  However, there may
be circumstances in which status is purely a subjective phenomenon,
lacking any objective correlate.  (People inside the community may
\emph{believe} that higher status is correlated with more expertise
but this would be an illusion.)

\subsubsection{Clustered Interaction}

Most diffusions occurs in communities that exhibit biased
interactions.  In general, executives in the same industry are more
likely to interact with each other than with executives in a
different industry.  School superintendents in the same state are
more likely to encounter each other than are superintendents in
different states. (At the limit---if these subcommunities were
completely sealed off from each other---then our model applies as it
is to each subcommunity.)\footnote{For a study of the influence of
social structure on opinion formation, see e.g.~\cite{wu:05,
sood:05}.}

\subsection{Leakage of Information}

Here, $p$ would be a function of time.  Typically we would expect
that $p$ would increase over time, as information about the
technology (by the already-converted or by third parties) leaks out.
As already noted, the  baseline diffusion process (stationary $p$)
is robust: small positive shocks to $p$ will on average have small
effects on the fraction of the population that adopts option A.

\subsection{Waves of Innovations}

Today it seems that the business world never rests.  No sooner has
one innovation passed from the scene---or at least from public
attention---than another one appears.  Becauase there are good
reasons for this---it is not accidental \cite{abrahamson:96}---one
can expect this pattern to continue indefinitely.

Although waves of innovation will obviously produce some new
patterns, we suspect that some of the present paper's results will
continue to hold.  In particular, we conjecture that if the waves
are composed of soft technologies, with $p$ around $1/2$ and strong
propensities to imitate, then the process will continue to be
Polya-like in two senses.  First, objectively uncertainty will be
very great: many adoption patterns will be possible ex ante. Second,
if the waves do not come too often then sample paths of diffusion
will settle down. Hence, subjectively it will feel as if the process
is moving toward a predetermined equilibrium---an illusion.

\section{Conclusion}

Because they diffuse with little objective information about their
effects, soft technologies pose challenges for decision makers.
Psychologists have long argued that when faced with such choice
problems, people use a reasonable though imperfect imitation
heuristic.  We have presented a mathematical model of diffusion that
combines this heuristic with agents' efforts to make
factually-grounded decisions and we established both analytically
and computationally that such processes exhibit clear stochastic
properties. We then showed that the dynamics of the model leads to
outcomes that appear to be deterministic in spite of being governed
by a stochastic process. In other words, when the objective evidence
for the adoption of a soft-technology is weak, any sample path of
this process quickly settles down to a fraction of adopters that is
not predetermined by the initial conditions: ex ante, every outcome
is just as (un)likely as every other.  When the objective evidence
is strong, the process settles down to a value that is determined by
the joint effect of the quality of the evidence and the agents'
competence. In neither case does the proportion of adopters settle
into either zero or one: pure herding does not occur except in
parametrically extreme situations.

Further, unlike informational cascades generated by fully rational
actors, the  process of the present model is robust:  diffusions
that have for a long time tilted massively toward one option cannot
be suddenly derailed by small infusions of new public information.
The fragility of cascades generated by fully Bayesian agents is, we
believe, an artifact of unrealistic assumptions of
hyper-rationality.  Diffusions may be \emph{initially} volatile, as
they are in the present model, but we believe that these processes
stabilize once the weight of public opinion has been brought to
bear.\footnote{Of course, the introduction of a new innovation can,
by restarting the process, destabilize it.   But that is not what is
producing fragility in the full-rationality cascades: these are not
robust against small shocks associated with the \emph{pre-existing}
options.  Moreover, as indicated earlier, we believe that the
present model can be extended to accommodate waves of innovation.}

\pagebreak

\bibliographystyle{alpha}
\bibliography{fads}

\pagebreak

\section*{Appendix}
\setcounter{prop}{1}

Consider the following variation of Polya's urn. There are two types
of alternatives, A and B. At time $t=0$ there are $s$ agents who
have already made their choices, of which $m_0$ have chosen A and
$n_0$ have chosen B. Denote the initial fraction of A-adherents to
be $f_0=m_0/s$. Define the dynamics recursively as follows. Let
$(M_t, N_t)$ be the (random) number of agents who have chosen A by
the end of period $t$. In period $t+1$, an agent chooses A with
probability
\begin{equation}
P_t=\alpha p+(1-\alpha) \frac{M_t}{M_t+N_t},
\end{equation}
where $\alpha, p \in [0,1]$, and chooses B with probability $1-P_t$.
If $\alpha=0$ this becomes a standard Polya's urn process. In the
rest of this appendix we will assume $\alpha>0$.

We are interested in studying the (random) fraction of agents who
have chosen A up to period $t$:
\begin{equation}
F_t=\frac{M_t}{M_t+N_t}.
\end{equation}
Conditional on the information at time $t$, with probability $P_t$,
\begin{equation}
\Delta F_{t+1} = F_{t+1}-F_t =
\frac{M_t+1}{M_t+N_t+1}-\frac{M_t}{M_t+N_t}=\frac{1-F_t}{t+s+1},
\end{equation}
while with probability $1-P_t$,
\begin{equation}
\Delta F_{t+1} =
\frac{M_t}{M_t+N_t+1}-\frac{M_t}{M_t+N_t}=\frac{-F_t}{t+s+1}.
\end{equation}
Hence
\begin{equation}
E_t[\Delta F_{t+1}]=\frac{P_t-F_t}{t+s+1} =
\frac{\alpha(p-F_t)}{t+s+1}.
\end{equation}
Separating out the mean term, we can write
\begin{equation}
\label{eq:dyn_F} \Delta F_{t+1}= \frac{\alpha(p-F_t)}{t+s+1} +
\frac{X_{t+1}}{t+s+1},
\end{equation}
where $X_t$ is a martingale difference with the conditional
distribution
\begin{equation}
X_{t+1}=
\begin{cases}
1-P_t & \text{with probability } P_t,\\
-P_t & \text{with probability } 1-P_t.
\end{cases}
\end{equation}

Eq.~(\ref{eq:dyn_F}) falls into the general class of
\emph{stochastic approximation} processes that have been studied
intensively in the statistics literature \cite{robbins:51,
pemantle:01}.

In what follows it is convenient to introduce an auxiliary
random variable
\begin{equation}
G_t=F_t-p,
\end{equation}
so that Eq.~(\ref{eq:dyn_F}) can be written as
\begin{equation}
\label{eq:dyn_G} \Delta G_{t+1}=\frac{-\alpha G_t+X_{t+1}}{t+s+1}.
\end{equation}

\begin{prop}
$EF_t\to p$ monotonously as $t\to \infty$.
\end{prop}
\begin{proof}
From Eq.~(\ref{eq:dyn_G}) we have
\begin{equation}
\label{eq:recursion_G} G_{t+1} = \left( 1-\frac\alpha{t+s+1} \right)
G_t + \frac{X_{t+1}}{t+s+1}.
\end{equation}
Taking expectation of both sides, we have
\begin{equation}
\label{eq:EG} EG_{t+1} = \left( 1-\frac\alpha{t+s+1} \right) EG_t =
\prod_{k=0}^t \left( 1-\frac\alpha{k+s+1} \right) g_0.
\end{equation}
Because
\begin{equation}
\sum_{k=0}^\infty \frac\alpha{k+s+1} = \infty,
\end{equation}
the infinite product in Eq.~(\ref{eq:EG}) converges to zero
monotonously. Thus $EG_t\to 0$ monotonously, or $EF_t\to p$
monotonously.
\end{proof}

\begin{lemma}
\label{lm:convergence} Assume $\alpha>0$. If
\begin{equation}
y_n=x_n+ \alpha \sum_{k=1}^{n-1} \frac{x_k}k
\end{equation}
converges, then $x_n\to 0$.
\end{lemma}
\begin{proof}
Suppose $x_n$ does not converge to 0. Then without loss of
generality we can assume that $x_n>\epsilon>0$ infinitely often. It
must also be that $x_n<\epsilon/2$ infinitely often, otherwise we
would have $x_n \ge \epsilon/2$ eventually and $y_n$ would diverge.
Because $y_n$ is Cauchy, we can find $N$ such that
$|y_m-y_n|<\epsilon/4$ for all $m,n\ge N$. Pick $n'\ge \max\{N,
2\alpha\}$ such that $x_{n'}<\epsilon/2$. Pick $m>n'$ such that
$x_m>\epsilon$. Let $n$ be the largest integer such that $n'\le n<m$
and $x_n<\epsilon/2$. It is clear that $m,n$ chosen this way satisfy
the following conditions:
\begin{equation}
\begin{cases}
m>n\ge N,\\
x_n<\epsilon/2, \quad x_m>\epsilon,\\
n<i<m \Rightarrow x_i \ge \epsilon/2 >0.
\end{cases}
\end{equation}
Now we have
\begin{multline}
x_m-x_n = y_m-y_n -\alpha \sum_{k=n}^{m-1} \frac{x_k}k \le y_m-y_n -
\alpha \frac{x_n}n\\ < y_m-y_n+\frac\alpha n(x_m-x_n) <
\frac\epsilon 4+ \frac 12 (x_m-x_n).
\end{multline}
Hence $x_m-x_n<\epsilon/2$. But this contradicts with the fact that
$x_n<\epsilon/2$ and $x_m>\epsilon$.
\end{proof}

\begin{prop}
$F_t\to p$ a.s. as $t\to\infty$.
\end{prop}
\begin{proof}
Replacing $t$ by $k$ and summing up $k=0, \dots, t-1$ in
Eq.~(\ref{eq:dyn_G}), we find a martingale:
\begin{equation}
\label{eq:tildeG} \tilde G_t = G_t - g_0 +\sum_{k=0}^{t-1}
\frac{\alpha G_k}{k+s+1} = \sum_{k=0}^{t-1} \frac{X_{k+1}}{k+s+1}.
\end{equation}
Furthermore, this martingale is nonnegative. The martingale
convergence theorem thus ensures that $\tilde G_t$ converges to a
(random) limit $\tilde G_\infty$ with probability 1. From
Eq.~(\ref{eq:tildeG}) and Lemma \ref{lm:convergence}, the almost
sure convergence of $\tilde G_t$ implies that $G_t\to 0$ almost
surely, or $F_t\to p$ almost surely.
\end{proof}

\begin{prop}
A higher value of $p$ yields a distribution of $F_t$ that
stochastically dominates a distribution of $F_t$ produced by a lower
value of $p$, for all $t>0$.
\end{prop}
\begin{proof}
Suppose $p'>p$. It suffices to show by induction that for all $t$,
$M_t'$ stochastically dominates $M_t$, or $P[M_t'> m]\ge P[M_t> m]$
for all $m\in\mathbb N$. The statement is correct for $t=0$ because
$m_0'=m_0$. Suppose the statement is correct for $t$, we prove for
$t+1$.

\emph{Case 1.} \quad $P[M_t'=m]\ge P[M_t=m]$.
\begin{eqnarray}
& & P[M_{t+1}'> m] \nonumber\\
&=&P[M_t'> m]+P[M_t'=m] \left[ \alpha
p'+(1-\alpha)\frac m{t+s} \right] \nonumber\\
&\ge& P[M_t> m]+P[M_t=m] \left[ \alpha p+(1-\alpha)\frac m{t+s}
\right] \nonumber\\
&=& P[M_t>m].
\end{eqnarray}

\emph{Case 2.} \quad $P[M_t'=m]< P[M_t=m]$.
\begin{eqnarray}
& & P[M_{t+1}'\le m] \nonumber\\
&=&P[M_t'< m]+P[M_t'=m] \left[ \alpha (1-p')+(1-\alpha) \left(
1-\frac m{t+s} \right) \right] \nonumber\\
&\le& P[M_t< m]+P[M_t=m] \left[ \alpha (1-p)+(1-\alpha) \left(
1-\frac m{t+s} \right)\right] \nonumber\\
&=& P[M_t\le m].
\end{eqnarray}
\end{proof}

\begin{prop}
\mbox{}
\begin{itemize}
\item[(i)] If $f_0<p$, then $EF_t$ is increasing in $\alpha$, for all
$t>0$.
\item[(ii)] If $f_0>p$, then $EF_t$ is decreasing in $\alpha$, for all
$t>0$.
\end{itemize}
\end{prop}
\begin{proof}
See Eq.~(\ref{eq:EG}).
\end{proof}

\begin{prop}
$\partial(\var F_t)/\partial \alpha <0$ for all $t>0$ iff either
\begin{itemize}
\item[(i)] $f_0<p$ and $p_0=(1-\alpha)f_0+\alpha p>1/2$, or
\item[(ii)] $f_0>p$ and $p_0=(1-\alpha)f_0+\alpha p<1/2$.
\end{itemize}
\end{prop}
\begin{proof}
``If'' part. \quad Taking variance of both sides of
Eq.~(\ref{eq:recursion_G}), we have
\begin{equation}
\label{eq:varG} \var G_{t+1} = \left( 1-\frac \alpha{t+s+1}
\right)^2 \var G_t + \frac 1{(t+s+1)^2} \var X_{t+1}.
\end{equation}
The variance of $X_{t+1}$ can be calculated as follows. First note
that
\begin{equation}
P_t = \alpha p + (1-\alpha) F_t = p+(1-\alpha)G_t.
\end{equation}
Then we can write
\begin{eqnarray}
\var X_{t+1} &=& E[X_{t+1}^2] = E[E_t[X_{t+1}^2]]=E[P_t(1-P_t)] \nonumber\\
&=& EP_t(1-EP_t) - \var P_t \nonumber\\
&=& EP_t(1-EP_t) - (1-\alpha)^2 \var G_t.
\end{eqnarray}
Plugging this back into Eq.~(\ref{eq:varG}), we obtain a recursive
relation for $\var G_t$:
\begin{equation}
\label{eq:rec_varG}\var G_{t+1} =
\frac{(t+s+2-2\alpha)(t+s)}{(t+s+1)^2} \var G_t +
\frac{EP_t(1-EP_t)}{(t+s+1)^2},
\end{equation}
where
\begin{equation}
\label{eq:Pt} EP_t=p+(1-\alpha)EG_t.
\end{equation}
It is clear from this recursive relation that the conclusion holds
if
\begin{equation}
\frac{\partial(EP_t(1-EP_t))}{\partial\alpha} <0.
\end{equation}

\emph{Case (i)} \quad By hypothesis $g_0<0$. From Eq.~(\ref{eq:EG})
we see that $EG_t$ (being a negative sequence) is increasing in both
$t$ and $\alpha$, so is $EP_t$ by Eq.~(\ref{eq:Pt}). Therefore
$EP_t>p_0>1/2$. Note that the function $f(x)=x(1-x)$ is decreasing
in $x$ when $x>1/2$. Thus $EP_t(1-EP_t)$ is decreasing in $\alpha$
for all $t$.

\emph{Case (ii)} \quad Similar to (i).

\medskip\noindent
``Only if'' part. \quad Letting $t=0$ in Eq.~(\ref{eq:rec_varG}), we
have
\begin{equation}
\var G_1 = \frac{p_0(1-p_0)}{(s+1)^2}.
\end{equation}
Taking partial derivative with respect to $\alpha$, we obtain
\begin{equation}
\frac{\partial(\var G_1)}{\partial \alpha} = \frac{(f_0-p)
(2p_0-1)}{(s+1)^2}.
\end{equation}
Thus, in order to have $\partial(\var F_1)/\partial \alpha <0$, we
must have $(f_0-p)(2p_0-1)<0$, i.e., either (i) or (ii).
\end{proof}

\begin{prop}
\mbox{}
\begin{itemize}
\item[(i)] $\partial(\var F_t)/\partial p < 0$ for all $t>0$ if $p\ge
1/2$ and $p_0>1/2$;
\item[(ii)] $\partial(\var F_t)/\partial p > 0$ for all $t>0$ if $p\le
1/2$ and $p_0<1/2$.
\end{itemize}
\end{prop}
\begin{proof}
We see from Eq.~(\ref{eq:Pt}) and Eq.~(\ref{eq:EG}) that
\begin{equation}
\frac{\partial(EP_t)}{\partial p}=1+(1-\alpha) \frac{\partial
(EG_t)}{\partial p} = 1 -(1-\alpha) \prod_{k=0}^{t-1} \left(
1-\frac\alpha{k+s+1} \right) >0.
\end{equation}
Taking partial derivative of $p$ on both sides of
Eq.~(\ref{eq:rec_varG}) yields
\begin{equation}
\label{eq:varG_p}\frac{\partial(\var G_{t+1})}{\partial p} =
\frac{(t+s+2-2\alpha)(t+s)}{(t+s+1)^2} \frac{\partial(\var
G_t)}{\partial p} + \frac{1-2EP_t}{(t+s+1)^2} \frac{\partial
(EP_t)}{\partial p}.
\end{equation}
The result now follows from the facts that (i) $EP_t>1/2$ for all
$t>0$ if $p\ge 1/2$ and $p_0>1/2$, and (ii) $EP_t<1/2$ for all $t>0$
if $p\le 1/2$ and $p_0<1/2$.
\end{proof}

\begin{prop}
Suppose $F_t'$ is a bigger process than $F_t$ in that $m_0'=km_0$
and $n_0'=kn_0$, where $k$ is an integer larger than one. In all
other respects the two processes are identical.
\begin{itemize}
\item[(i)] $EF_t'<EF_t$ for all $t>0$ iff $f_0<p$.
\item[(ii)] $EF_t'>EF_t$ for all $t>0$ iff $f_0>p$.
\end{itemize}
\end{prop}
\begin{proof}
When $f_0<p$ we have $g_0<0$. In this case Eq.~(\ref{eq:EG}) shows
that $EG_t$ is decreasing in $s$. Hence $EF_t'<EF_t$. When $f_0>p$
we have $EF_t'>EF_t$ similarly.
\end{proof}

\begin{prop}
All else equal, the distribution of $F_t$ is stochastically
increasing, in a strong sense, in $f_0$, for all $t>0$.
\end{prop}
\begin{proof}
We only need to show that $M_t$ is stochastically increasing in
$f_0$ in a strong sense. We prove this by induction. The proposition
obviously holds for $t=1$. Suppose $t$ is correct, we prove for
$t+1$. Let $M_t$ and $M_t'$ be the two stochastic processes
generated by $f_0$ and $f_0'$, assuming that $f_0<f_0'$. By
induction hypothesis, there exist $s\le a<b<c<d\le t+s$ such that
$P[M_t'=i]<P[M_t=i]$ for all integers $i\in [a,b]$,
$P[M_t'=i]=P[M_t=i]$ for all integers $i\in [b+1,c-1]$ (possibly
none), and $P[M_t'=i]>P[M_t=i]$ for all integers $i\in [c,d]$.
Furthermore, $[a,d]$ covers the support of $M_t$ and $M_t'$.

For all $i=c+1, \dots, d+1$, we have
\begin{eqnarray}
\label{eq:strong_dominance} P[M_{t+1}'=i] &=& P[M_t'=i-1] \left[
\alpha p+(1-\alpha)\frac
{i-1}{t+s} \right] \nonumber\\
& & \mbox{}+ P[M_t'=i] \left[ \alpha(1-p)+(1-\alpha)\left( 1-\frac
i{t+s}
\right) \right] \nonumber\\
&>& P[M_t=i-1] \left[ \alpha p+(1-\alpha)\frac {i-1}{t+s} \right]
\nonumber\\
& & \mbox{}+ P[M_t=i] \left[ \alpha(1-p)+(1-\alpha)\left( 1-\frac
i{t+s}
\right) \right] \nonumber\\
&=& P[M_{t+1}=i].
\end{eqnarray}
Similarly, for all $i=a, \dots, b$, we have
$P[M_{t+1}'=i]<P[M_{t+1}=i]$.

To study the relationship between the probability atoms of
$M_{t+1}'$ and $M_{t+1}$ at the ``boundaries'', distinguish the
following two cases:

\emph{Case 1.} \quad There are no intermediate ``equal'' states
($b+1=c$).

Whatever the relationship between $P[M_{t+1}'=c]$ and
$P[M_{t+1}=c]$, we have $M_{t+1}'$ strongly dominates $M_{t+1}$.

\emph{Case 2.} \quad There exist some intermediate ``equal'' states
($b+1<c$).

It can be checked, in a fashion similar to
Eq.~(\ref{eq:strong_dominance}), that (a) $P[M_{t+1}'=b+1]\le
P[M_{t+1}=b+1]$, (b) $P[M_{t+1}'=c]\ge P[M_{t+1}=c]$, and (c)
$P[M_{t+1}'=i]=P[M_{t+1}=i]$ for all $i=b+2, \dots, c-1$ (if any).
Hence $M_{t+1}'$ strongly dominates $M_{t+1}$.
\end{proof}

\begin{prop}
$\partial (\var F_t)/\partial f_0 <0$ for all $t>0$ if $p_0>1/2$ and
$p>1/2$.
\end{prop}
\begin{proof}
Using Eq.~(\ref{eq:Pt}) and Eq.~(\ref{eq:EG}), it can be calculated
that
\begin{equation}
\frac{\partial(EP_t(1-EP_t))}{\partial f_0} = (1-2EP_t)(1-\alpha)
\prod_{k=0}^{t-1} \left( 1-\frac\alpha{k+s+1} \right).
\end{equation}
If $p_0>1/2$ and $p>1/2$, then because $EP_t$ converges to $p$
monotonously, we have $EP_t>1/2$ for all $t>0$, and therefore
\begin{equation}
\frac{\partial(EP_t(1-EP_t))}{\partial f_0}<0
\end{equation}
for all $t>0$. The result now follows from Eq.~(\ref{eq:rec_varG}).
\end{proof}

\begin{prop}
\mbox{}
\begin{itemize}
\item[(i)] If $1/2<\alpha\le 1$, then as $t\to \infty$, $\sqrt t(F_t-p)$
converges in distribution to a normal distribution with mean zero
and variance $p(1-p)/(2\alpha-1)$. In particular, $\var F_t =
O(1/t)$.
\item[(ii)] If $0< \alpha \le 1/2$, then $\var F_t \to 0$
slower than $O(1/t)$.
\end{itemize}
\end{prop}
\begin{proof} To understand this proposition, define $v_t=(t+s)\var G_t$ and rewrite
Eq.~(\ref{eq:rec_varG}) as
\begin{equation}
\label{eq:vt} v_{t+1} - v_t =
\frac{EP_t(1-EP_t)-(2\alpha-1)v_t}{t+s+1}.
\end{equation}
When $t$ is large, we have
\begin{equation}
\Delta v_{t+1} \sim \frac{p(1-p)-(2\alpha-1)v_t}t,
\end{equation}
Hence if $2\alpha-1>0$, $v_t$ is dragged to the limit
$p(1-p)/(2\alpha-1)$.

It is also clear from Eq.~(\ref{eq:vt}) that when $0<\alpha<1/2$ we
have $v_t\to\infty$ in general, so $\var G_t$ converges slower than
$O(1/t)$.

A rigorous proof for (i), however, is too technical to be presented
here. Various authors have given proofs for general stochastic
approximation processes. For references see \cite{blum:54,
chung:54}.
\end{proof}

\end{document}